\documentclass[aps,prb,reprint,superscriptaddress,twocolumn,showpacs,floatfix]{revtex4-1}

\usepackage{graphicx}
\usepackage{amsfonts,amsmath,amssymb}
\usepackage{url}
\usepackage[utf8]{inputenc}
\usepackage{hyperref}
\hypersetup{colorlinks=false,pdfborder={0 0 0},}

\newcommand{\siesta}{\textsc{siesta}}

\begin{document}

\title{Ab initio calculation of energy levels for phosphorus donors in silicon}

\author{J.~S.~Smith}
\affiliation{Chemical and Quantum Physics, School of Science, RMIT University, Melbourne VIC 3001, Australia}

\author{A.~Budi}
\affiliation{Materials Chemistry, Nano-Science Center, Department of Chemistry, University of Copenhagen, Universitetsparken 5, 2100 K{\o}benhavn {\O}, Denmark}

\author{M.~C.~Per}
\affiliation{Data 61 CSIRO, Door 34 Goods Shed, Village Street, Docklands VIC 3008, Australia}
    
\author{N.~Vogt}
\affiliation{Chemical and Quantum Physics, School of Science, RMIT University, Melbourne VIC 3001, Australia}

\author{D.~W.~Drumm}
\affiliation{Chemical and Quantum Physics, School of Science, RMIT University, Melbourne VIC 3001, Australia}
\affiliation{Australian Research Council Centre of Excellence for Nanoscale BioPhotonics, School of Science, RMIT University, Melbourne, VIC 3001, Australia}

\author{L.~C.~L.~Hollenberg}
\affiliation{Centre for Quantum Computation and Communication Technology, School of Physics, The University of Melbourne, Parkville, 3010 Victoria, Australia}

\author{J.~H.~Cole}
\affiliation{Chemical and Quantum Physics, School of Science, RMIT University, Melbourne VIC 3001, Australia}

\author{S.~P.~Russo}
\affiliation{Chemical and Quantum Physics, School of Science, RMIT University, Melbourne VIC 3001, Australia}

\begin{abstract}

The $s$ manifold energy levels for phosphorus donors in silicon are important input parameters for the design and modelling of electronic devices on the nanoscale.
In this paper we calculate these energy levels from first principles using density functional theory.
The wavefunction of the donor electron's ground state is found to have a form that is similar to an atomic $s$ orbital, with an effective Bohr radius of 1.8~nm.
The corresponding binding energy of this state is found to be 41~meV, which is in good agreement with the currently accepted value of 45.59~meV.
We also calculate the energies of the excited $1s(T_{2})$ and $1s(E)$ states, finding them to be 32 and 31~meV respectively.
These results constitute the first ab initio confirmation of the $s$ manifold energy levels for phosphorus donors in silicon.

\end{abstract}

\maketitle

Phosphorus donors in silicon have long been important for electronic devices but are now seen as central to the development of silicon based quantum information processing~\cite{Kane1998a,Morello2010a,Fuechsle2012a,Pla2012a,Hill2015a}.
The phosphorus donor electron has been shown to have long spin coherence times in the laboratory, which make these donors excellent candidates for spin qubits~\cite{Morello2010a,Pla2012a}.
Moreover, during the last decade a technique of phosphorus $\delta$~doping, based on scanning tunneling microscope (STM) lithography, has led to a variety of new electronic devices in silicon~\cite{Zwanenburg2013a}.
This $\delta$~doping technique has been used to make a quantum dot of seven donors~\cite{Fuechsle2010a} and a transistor with a gate island that consists of only one phosphorus donor~\cite{Fuechsle2012a}.
Another novel electronic device is the quantized electron pump of \citet{Tettamanzi2014a}, which demonstrates charge pumping of single electrons through a phosphorus donor.
Finally, the wavefunction of the donor electron has even recently been imaged using an STM~\cite{Salfi2014a}.
These images have been analysed using tight binding~\cite{Usman2016a} and effective mass theory~\cite{Saraiva2016a}; whilst the latter provides a qualitative description, the tight binding method is precise enough to pinpoint the atomic position of a single phosphorus donor in the silicon lattice.
Although semi-empirical approaches have successfully been used to model the properties of these donor devices, a full ab initio treatment of the electronic structure of these donors has to-date not been possible.
Here we present such a treatment.

At low doping densities it is well known that the phosphorus donor electrons occupy the lowest energy conduction band of silicon.
In bulk silicon this band is sixfold degenerate but the degeneracy is lifted by a valley splitting when silicon is doped~\cite{Kohn1955a, Mayur1993a}, resulting in three nondegenerate states.
These states are, in order of increasing energy, a singlet [$1s(A_{1})$], a triplet [$1s(T_{2})$], and a doublet [$1s(E)$]~\cite{Aggarwal1965b}.
Only the ground state [$1s(A_{1})$] is populated~\cite{Aggarwal1965b} at liquid helium temperatures ($\sim4~\mathrm{K}$), whereas at higher temperatures ($\geq30$~K) the populations of the excited $1s(T_{2})$ and $1s(E)$ states become observable due to thermal broadening~\cite{Aggarwal1964a,Mayur1993a}.

Over a decade ago, theoretical methods for describing point defects in semiconductors were separable into two categories: ``methods for deep defects and methods for shallow defects: the former defect class is treated by ab initio methods, ... while for the latter class approximate one-electron theories ... are used''~\cite{Overhof2004a}.
Traditionally, shallow defects in silicon like phosphorus donors could not be treated by ab initio methods because the wavefunctions of such defects are partially delocalized.
Today, however, this statement does not hold true, as in the last ten years innovations in modern computing technologies have made much larger computational resources available to scientific research.
Recently it has been shown that shallow defects are now within the reach of ab initio methods such as density functional theory (DFT)~\cite{Yamamoto2009a}.

In this paper we calculate the energies of the $s$ manifold states [$1s(A_{1})$, $1s(T_{2})$, and $1s(E)$] of a phosphorus donor electron in silicon from first principles using DFT.
We also compute the wavefunction of the donor electron's ground state [$1s(A_{1})$].\
From this we estimate the effective Bohr radius of the electron by fitting to this wavefunction.
We find DFT significantly underestimates the energies of the $s$ manifold states.
This is a known problem and, as will be discussed, we correct these energies using the ground state wavefunction, via the method described in Ref.~\onlinecite{Yamamoto2009a}.
In this way we are able to obtain ionisation energies for the donor electron that are in good agreement with the currently accepted values.
To the best of our knowledge these results are the first ab initio confirmation of the $s$ manifold energy levels for a phosphorus donor in silicon.

The Lyman spectrum for Group V donors in silicon was first measured by Aggarwal et al.\ in 1965~\cite{Aggarwal1965a}.
These measurements do not give the binding energy of the donor electrons but rather the energy splitting between the ground and excited states; namely, the energy splitting between the $1s(A_{1})$ and $3p_{\pm}$ states.
The binding energy of the phosphorus donor electron that is reported in Ref.~\onlinecite{Aggarwal1965a} was computed by ``adding the theoretically calculated binding energy of 2.90~meV for the $3p_{\pm}$ state~\cite{Kohn1955a} to the energy of the transition $1s(A_{1}) \to 3p_{\pm}$''.
The binding energy of the phosphorus donor electron was thereby found to be 45.31~meV~\cite{Aggarwal1965a,Jagannath1981a}.

In 1969, Faulkner used effective mass theory (EMT) to calculate the energy levels of the ground and excited states of a donor electron for Group V donor atoms in silicon~\cite{Faulkner1969a}.
For the phosphorus donor electron the binding energies of the $3p_{\pm}$ and $1s(A_{1})$ states were found to be 3.12~meV and 31.27~meV, respectively~\cite{Faulkner1969a}.
The theoretically calculated binding energy of the excited $3p_{\pm}$ state is in good agreement with experiment, whereas the binding energy of the $1s(A_{1})$ state is not~\cite{Faulkner1969a}.
Later, in 1981, using the theoretical correction of \citet{Faulkner1969a} and a new experimental technique that produced narrower linewidths in the excitation spectra, \citet{Jagannath1981a} reported a binding energy of 45.59~meV for the phosphorus donor electron.

More recently it has been demonstrated that EMT, with effective potentials calculated from ab initio methods,  is capable of reproducing the accepted values for the binding energies of the $s$ manifold states~\cite{Klymenko2015a}.
In addition, a model for a phosphorus donor in silicon that goes ``beyond effective mass theory'' has been introduced~\cite{Wellard2005a}.
In Ref.~\onlinecite{Wellard2005a} the binding energy was used as a fitting parameter together with non-static screening effects in a model that provided an excellent account of the $s$ manifold of states.
This study shows that the binding energy is also an important quantity for theoretical modelling.
The same fact is highlighted by Ref.~\onlinecite{Rahman2007a}, where the hyperfine Stark effect is investigated using a truncated Coulomb potential to approximate the impurity potential of an ionized phosphorus donor~\cite{Rahman2007a}.
The truncation of the Coulomb potential was found by adjusting a free parameter ``to obtain the experimental ground state energy of 45.6~meV''~\cite{Rahman2007a}.

The binding energies of Group V donors in silicon have also been used as input parameters to modelling of the hyperfine Stark effect with EMT~\cite{Pica2014b}.
EMT has been shown to be capable of reproducing the wavefunction of a phosphorus donor electron that is predicted by tight binding theory~\cite{Gamble2015a}.
The results in Ref.~\onlinecite{Gamble2015a} were benchmarked against the currently accepted value for the binding energy of a phosphorus donor electron in silicon.
Knowledge of the binding energy, and specifically the valley splitting, was needed to choose the exact form of the central-cell corrections, i.e. a central cell with tetrahedral, rather than spherical, symmetry~\cite{Gamble2015a}.

\begin{table}[t!]
    \caption{\label{tab:table1}A list of the supercells that have been studied in this work, showing the number of atoms in each supercell and the real space dimensions of each of the cells (in units of simple-cubic unit cells). The dimensions of each simple-cubic unit cell are $0.546~\mathrm{nm} \times 0.546~\mathrm{nm} \times 0.546~\mathrm{nm}$.}
    \begin{ruledtabular}
        \begin{tabular}{c c}
            Number of atoms & Dimensions of supercell (unit cells) \\
            \hline
            216             & $3\times3\times3$    \\
            512             & $4\times4\times4$    \\
            1000            & $5\times5\times5$    \\
            1728            & $6\times6\times6$    \\
            2744            & $7\times7\times7$    \\
            4096            & $8\times8\times8$    \\
            5832            & $9\times9\times9$    \\
            8000            & $10\times10\times10$ \\
            10648          & $11\times11\times11$ \\
        \end{tabular}
    \end{ruledtabular}
\end{table}

The first large-scale atomic simulations performed on a Group V donor in silicon using DFT were those presented in Ref.~\onlinecite{Yamamoto2009a}.
In this study the electronic properties of an arsenic donor in silicon were calculated for systems that ranged in size from 512 to 10,648 atoms.
DFT has also been used to simulate phosphorus donors in silicon, with systems ranging in size from 54 to 432 atoms~\cite{Greenman2013a}.
However, as we will show, these latter system sizes are not large enough to isolate the phosphorus donor electron from its periodic images.
The confinement of the donor electron is thereby increased, which artificially raises the binding energy of the electron.
The binding energy of the phosphorus donor electron was therefore unable to be reported in Ref.~\onlinecite{Greenman2013a}.

In this paper we present the results of electronic structure calculations performed on a single phosphorus donor in silicon with DFT.
This approach has previously been benchmarked in a number of other studies~\cite{Carter2011a,Budi2012a,Carter2013a,Drumm2013b,Drumm2014a,Drumm2013a,Smith2015a}.
For more information on this method and its benchmarking see the supplemental material.
We have employed the \siesta~package~\cite{Soler2002a,Artacho2008a} to carry out calculations on systems that range in size up to 10,648 atoms.
Table~\ref{tab:table1} lists each of the supercell sizes that have been studied by number of atoms and the dimensions of the supercells in real space.
These calculations have been performed using periodic boundary conditions.

\begin{figure}[b!]
    \includegraphics{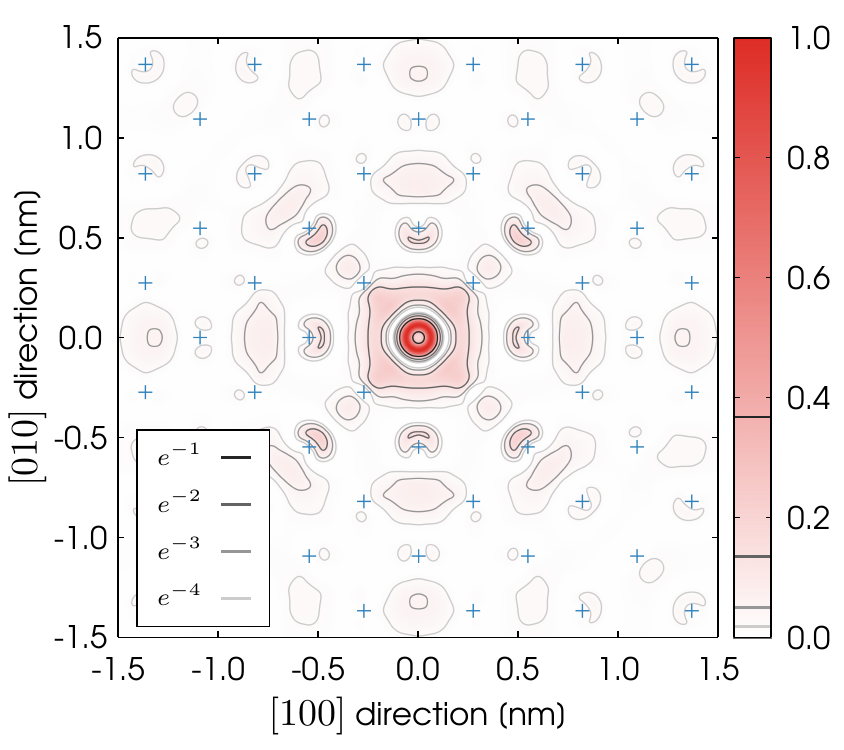}
    \caption{\label{fig:figure1}(Color online). A two dimensional slice of the probability density ($|\psi|^{2}$) for the phosphorus donor electron inside the dopant plane. The wavefunction of the donor electron has been calculated using a supercell of 10,648 atoms and the maximum of the probability density has been normalized to one. The contours show where this probability density is equal to a negative integer power of $e$. The blue pluses mark the positions of the in-plane silicon atoms.}
\end{figure}

We have calculated the wavefunction, $\psi$, of the donor electron's ground state for each of the supercells listed in Table~\ref{tab:table1}.
A two dimensional slice of the probability density, $|\psi|^{2}$, for the largest supercell studied in this work is plotted in Fig.~\ref{fig:figure1}.
This slice is computed by evaluating the wavefunction in the silicon (001) plane that contains the phosphorus donor.
The maximum of the probability density in this slice has been normalized to one.
In the (001) plane the majority of the probability density can be seen to be within $\sim0.5$~nm of the donor site, which is located at the origin in Fig.~\ref{fig:figure1}.
The wavefunction of the donor electron has a form that is similar to an atomic $s$ orbital.
The corresponding probability density can be seen to decay to approximately 2\% of its maximum value at a distance of $\pm1.5$~nm from the donor site in the $\left[100\right]$ and $\left[010\right]$ crystallographic directions.

The apparent hydrogenic character of the donor electron's wavefunction is compatible with an effective Bohr model of the electron.
Phosphorus is a shallow defect in silicon so it is reasonable to treat the wavefunction of the Kohn-Sham eigenvalue, calculated within DFT, as an independent single particle state that can be modelled by a simple exponential function.
The wavefunction of the donor electron can therefore be described by the envelope function ${F(r) = A \exp{(-r / a_{0}^{\ast})}}$ where $A$ is a normalisation constant and $a_{0}^{\ast}$ is an effective Bohr radius.
It is then possible to calculate the effective Bohr radius of the donor electron by fitting its wavefunction with this envelope function.
However, it is first necessary to spherically average the wavefunction of the donor electron because $F(r)$ is radially symmetric and $\psi$ is not.

\begin{figure}[t!]
    \includegraphics{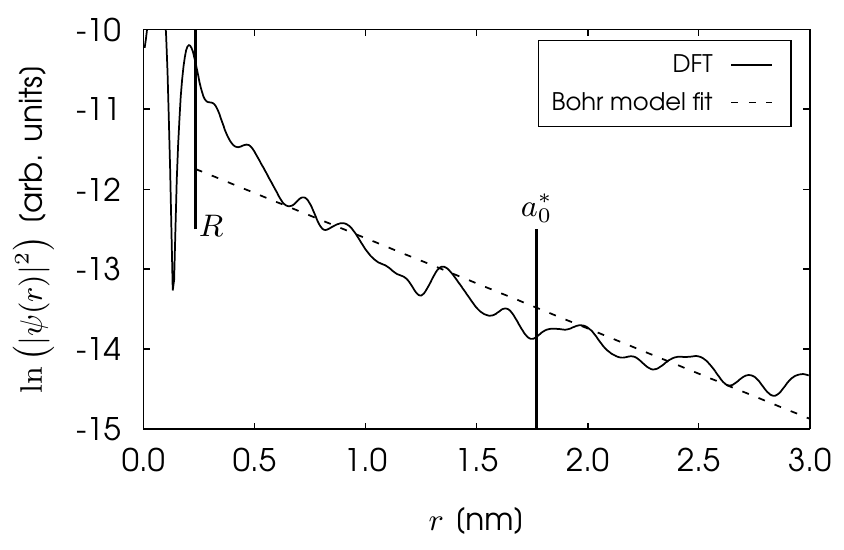}
    \caption{\label{fig:figure5}Natural logarithm of the spherically averaged probability density for the phosphorus donor electron [$\ln{\left(|\psi(r)|^{2}\right)}$] versus radial distance from the donor site [$r$] (solid line). A fit to this probability density described by the natural log of the square of the envelope function [$\ln{\left(F(r)^{2}\right)}$] (dashed line). The model radius [$R$] and effective Bohr radius [$a_{0}^{\ast}$] are also shown (vertical solid lines). The wavefunction of the donor electron has been calculated using a supercell of 10,648 atoms.}
\end{figure}

Figure~\ref{fig:figure5} shows the natural logarithm of the spherically averaged probability density for the phosphorus donor electron, $\ln{\left(|\psi(r)|^{2}\right)}$, plotted against radial distance from the donor site, $r$.
The domain in this figure includes the core region of the phosphorus atom, which in our model is described by a Troullier-Martins pseudopotential~\cite{Troullier1991a}.
A pseudopotential will deviate from a Coulombic potential in the core region.
The envelope function is not applicable within the core region because a hydrogenic wavefunction is not a valid solution here.
We have therefore fitted the wavefunction of the donor electron on the domain $[R,3.0]$~nm, where $R$ is termed the model radius.
The model radius must be chosen such that the effects of the core region on the wavefunction do not influence the accuracy of the exponential fit.
Nor can the model radius be so large that the whole of the wavefunction's exponential decay is not captured by the fit.
We have set the model radius equal to the atomic nearest neighbour distance, which has a value of 0.235~nm in silicon\cite{Kittel1954a}.
As can be seen from Fig.~\ref{fig:figure5}, this value for the model radius satisfies our two requirements.

In EMT, it is possible to derive two Bohr radii for the donor electron: one corresponds to the longitudinal effective mass, $m_{\parallel}$, of bulk silicon and the other to the transverse mass, $m_{\perp}$.
The geometric average of these two radii is given by $a_{\perp}^{2/3}a_{\parallel}^{1/3}$.
By fitting $\ln{\left(F(r)^{2}\right)}$ to $\ln{\left(|\psi(r)|^{2}\right)}$, we find the effective Bohr radius to be 1.8~nm.
This value is in good agreement with 2.087~nm, which is the geometric average of the two effective Bohr radii reported in Ref.~\onlinecite{Koiller2001a}.
By reconsidering Fig.~\ref{fig:figure1}, we can see that the effective Bohr radius can be thought of as the radial distance within which the vast majority of the probability density corresponding to the donor electron is contained.

\begin{figure}[b!]
    \includegraphics{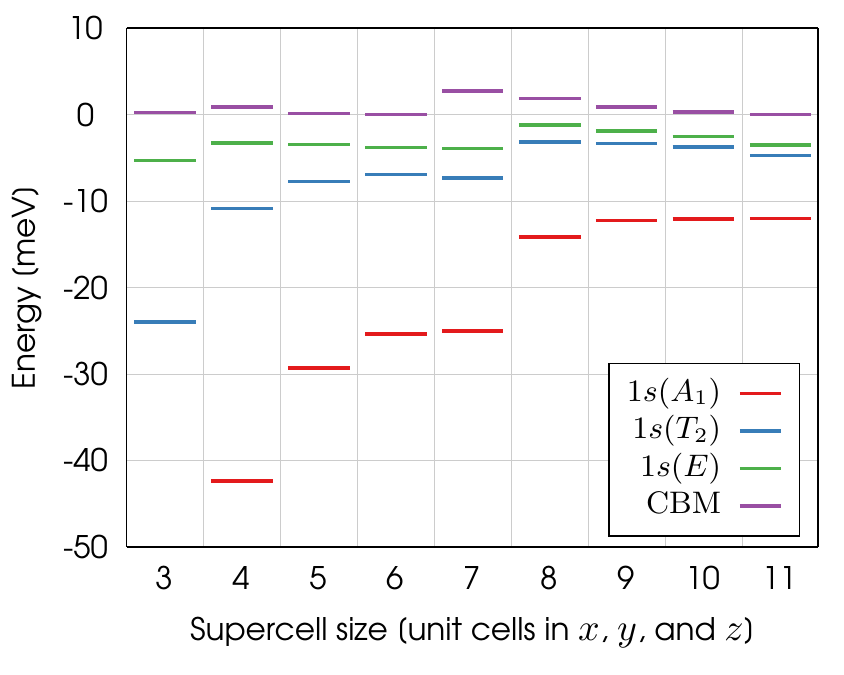}
	\caption{\label{fig:figure3}(Color online). The energy levels of the $1s(A_{1})$, $1s(T_{2})$, and $1s(E)$ states for supercells that range in size from 216 to 10,648 atoms. The energy of the conduction band minimum (CBM) of bulk silicon is also shown for each supercell size. The conduction band minimum of the supercell containing 10,648 atoms has been set to energy zero.}
\end{figure}

The $s$ manifold energy levels for the phosphorus donor electron are shown in Fig.~\ref{fig:figure3}.
These energies are plotted relative to the conduction band minimum of bulk silicon, calculated using a supercell of 10,648 atoms, which is set to energy zero in the figure.
Figure~\ref{fig:figure3} illustrates how the energies of the $1s(A_{1})$, $1s(T_{2})$, and $1s(E)$ states increase as the size of the supercell is increased.
We suggest the energy levels for the smaller supercells are artifically lowered due to the electron's interaction with its periodic images, which increases the confinement of the donor electron as the size of the supercell is decreased.
The energy levels are converged to within 1~meV for a supercell of 10,648 atoms.
These results justify the use of such a large supercell for the calculation of these energies.

The binding energies of the $1s(A_{1})$, $1s(T_{2})$, and $1s(E)$ states can be calculated for each supercell by taking the difference between the energy levels and the conduction band minimum of bulk silicon.
In Fig.~\ref{fig:figure3} the larger supercells significantly underestimate these energies.
This discrepancy is due to the fact the Kohn-Sham eigenvalues are single particle energies and do not correspond exactly to true excitations of the system.
This statement applies to energy levels that are unoccupied, as DFT is a ground state theory.
We therefore need another way to calculate the binding energies of the $s$ manifold energy levels.
This is provided by the method described in Ref.~\onlinecite{Yamamoto2009a}, where the binding energy of each state is calculated directly from its wavefunction.

The method of Ref.~\onlinecite{Yamamoto2009a} allows us to calculate the binding energies of the $1s(A_{1})$, $1s(T_{2})$, and $1s(E)$ states directly from their wavefunctions and the impurity potential of the phosphorus donor.
We begin by writing down the screened impurity potential of the phosphorus donor;~\cite{Yamamoto2009a}
\begin{equation} \label{eq:ScreenedImpurityPotential}
    V(\mathbf{r}) = \int_{-\infty}^{\infty} \epsilon^{-1}(q) V'(q) \exp{(-i \mathbf{q} \cdot \mathbf{r})} \frac{d^{3}q}{\left(2\pi\right)^{3}}
\end{equation}
where $V'(q)$ is the Fourier transform of the unscreened impurity potential. The dielectric screening is described by a nonlinear function~\cite{Nara1966a,Pantelides1974a,Wellard2005a};
\begin{equation} \label{eq:dielectricfunction}
    \epsilon^{-1}(q) = \frac{A q^{2}}{q^{2} + \alpha^{2}} + \frac{\left(1 - A\right) q^{2}}{q^{2} + \beta^{2}} + \frac{\gamma^{2}}{\epsilon(0)\left(q^{2} + \gamma^{2}\right)}
\end{equation}
with $A = 1.175$, $\alpha = 0.7572$, $\beta = 0.3123$, $\gamma = 2.044$, and the relative permittivity of silicon $\epsilon(0)=11.4$.
The constants $A$, $\alpha$, $\beta$, and $\gamma$ were found by fitting the above function to the $q$ dependent dielectric screening in silicon, which was calculated from the random phase approximation~\cite{Pantelides1974a}.
The kinetic and potential energies of the donor electron can then be computed;
\begin{equation}
    T = \frac{1}{2} \int_{-\infty}^{\infty} \psi^{*}(\mathbf{r}) \left(\frac{dV(\mathbf{r})}{d\mathbf{r}} \cdot \mathbf{r}\right) \psi(\mathbf{r}) d^{3}\mathbf{r}
\end{equation}
and
\begin{equation}
    U = \int_{-\infty}^{\infty} \psi^{*}(\mathbf{r}) V(\mathbf{r}) \psi(\mathbf{r}) d^{3}\mathbf{r}
\end{equation}
where $\psi$ is the wavefunction calculated from DFT.
Finally, we can calculate the binding energy of the donor electron:
\begin{equation}
    E = T + U
\end{equation}
For more information on this derivation, see the supplemental material.

\begin{table}[t!]
    \caption{\label{tab:table2}Binding energies for the $s$ manifold states of a phosphorus donor electron calculated using experiment, EMT, and DFT. The binding energies of DFT were calculated from wavefunctions computed with a supercell containing 10,648 atoms. Reference~\onlinecite{Jagannath1981a} uses a theoretical correction from Ref.~\onlinecite{Faulkner1969a}. Reference~\onlinecite{Mayur1993a} uses a theoretical correction from Ref.~\onlinecite{Faulkner1969a}. Reference~\onlinecite{Klymenko2015a} uses EMT with effective potentials calculated from ab initio methods. Reference~\onlinecite{Wellard2005a} is theory using the so-called band minima basis (BMB) method, in which the energy of the $1s(A_{1})$ state is fit to.}
    \begin{ruledtabular}
        \begin{tabular}{l c c c}
            & $1s(A_{1})$ & $1s(T_{2})$ & $1s(E)$ \\
            \hline
            Exp. \& EMT & 45.59~\cite{Jagannath1981a} & 33.88~\cite{Mayur1993a} & 32.54~\cite{Mayur1993a} \\
            EMT\cite{Klymenko2015a} & 45.40 & 33.86 & 32.08 \\
            BMB\cite{Wellard2005a} & 45.5 & 29.1 & 27.1 \\
            DFT (this work) & 41 & 32 & 31 \\
        \end{tabular}
    \end{ruledtabular}
\end{table}

Table~\ref{tab:table2} presents the binding energies of the $s$ manifold states calculated using experiment, EMT, and DFT.
The binding energy of the $1s(A_{1})$ state calculated using DFT (this work) is equal to 41~meV.
This energy is in good agreement with the accepted value of 45.59~meV, which has been calculated from the combination of an experimental measurement~\cite{Jagannath1981a} and a theoretical correction~\cite{Faulkner1969a}.
In addition, we find the binding energies of the excited $1s(T_{2})$ and $1s(E)$ states to be 32~meV and 31~meV, respectively.
These values are in excellent agreement with the other values listed in Table~\ref{tab:table2}, agreeing to within 2~meV.
The binding energies of the two excited states appear to be in better agreement with the accepted values for these energies than the energy of the donor electron's ground state.

In summary, we have calculated the wavefunction of a phosphorus donor electron in silicon with DFT.
This wavefunction is then used to compute the effective Bohr radius of the donor electron.
We employ a hydrogenic model of this electron and thereby find its Bohr radius to be 1.8~nm.
In addition, we compute the binding energy of the donor electron's ground state, which is found to be in good agreement with the currently accepted value.
The energies of the excited $1s(T_{2})$ and $1s(E)$ states are found to be in excellent agreement with the accepted values.
These results constitute the first ab initio calculation of the $s$ manifold energy levels for a single phosphorus donor in silicon.

This work was supported by computational resources provided by the Australian Government through the National Computational Infrastructure under the National Computational Merit Allocation Scheme. D.W.D. acknowledges the support of the ARC Centre of Excellence for Nanoscale BioPhotonics (CE140100003).

\appendix

\section{Supplemental information}

\subsection{\label{sec:methodology}Density functional theory}

The electronic structure calculations were performed with density functional theory (DFT) using the \siesta package~\cite{Soler2002a,Artacho2008a}.
We have employed the Perdew-Burke-Ernzerhof (PBE) exchange correlation (XC) functional in the generalised gradient approximation (GGA)~\cite{Perdew1996a}.
Application of the GGA to phosphorus-doped silicon systems in the past has produced results that are in good agreement with experiment~\cite{Liu2003a}.
The total energies of each of the supercells were converged to within 0.1~meV using a planewave energy cutoff of 300~Ry and a Fermi-Dirac occupation function at a temperature of 0~K.
Atomic potentials were described by norm-conserving Troullier-Martins pseudopotentials~\cite{Troullier1991a}.

We have variationally solved the Kohn-Sham equations using a basis set of localised atomic orbitals that was optimised for phosphorus-doped silicon using the simplex method~\cite{Budi2012a}.
The basis set was double-$\zeta$ polarised and was comprised of 13 radial functions.
In Ref.~\onlinecite{Drumm2013b}, localised single-$\zeta$ and double-$\zeta$ polarised bases, and a delocalised planewave basis were used to calculate the valley splitting for a phosphorus $\delta$ doped monolayer in silicon.
Despite the higher precision of the planewave basis, the double-$\zeta$ polarised basis was shown to ``[retain] the physics of the planewave description''~\cite{Drumm2013b}.

We relaxed the crystallographic structure of bulk silicon using this basis set and found the lattice constant to be 5.4575~\AA.
This value is in good agreement with the experimental value of 5.431~\AA~\cite{Becker1982a}.
The overestimation of the lattice constant by approximately 0.5\% is lower than the usual systematic deviation of the lattice constant that is expected from the PBE XC functional, which is a 1\% deviation.

\subsection{\label{sec:benchmarking}Benchmarking of density functional theory}

\begin{figure}[t!]
    \includegraphics{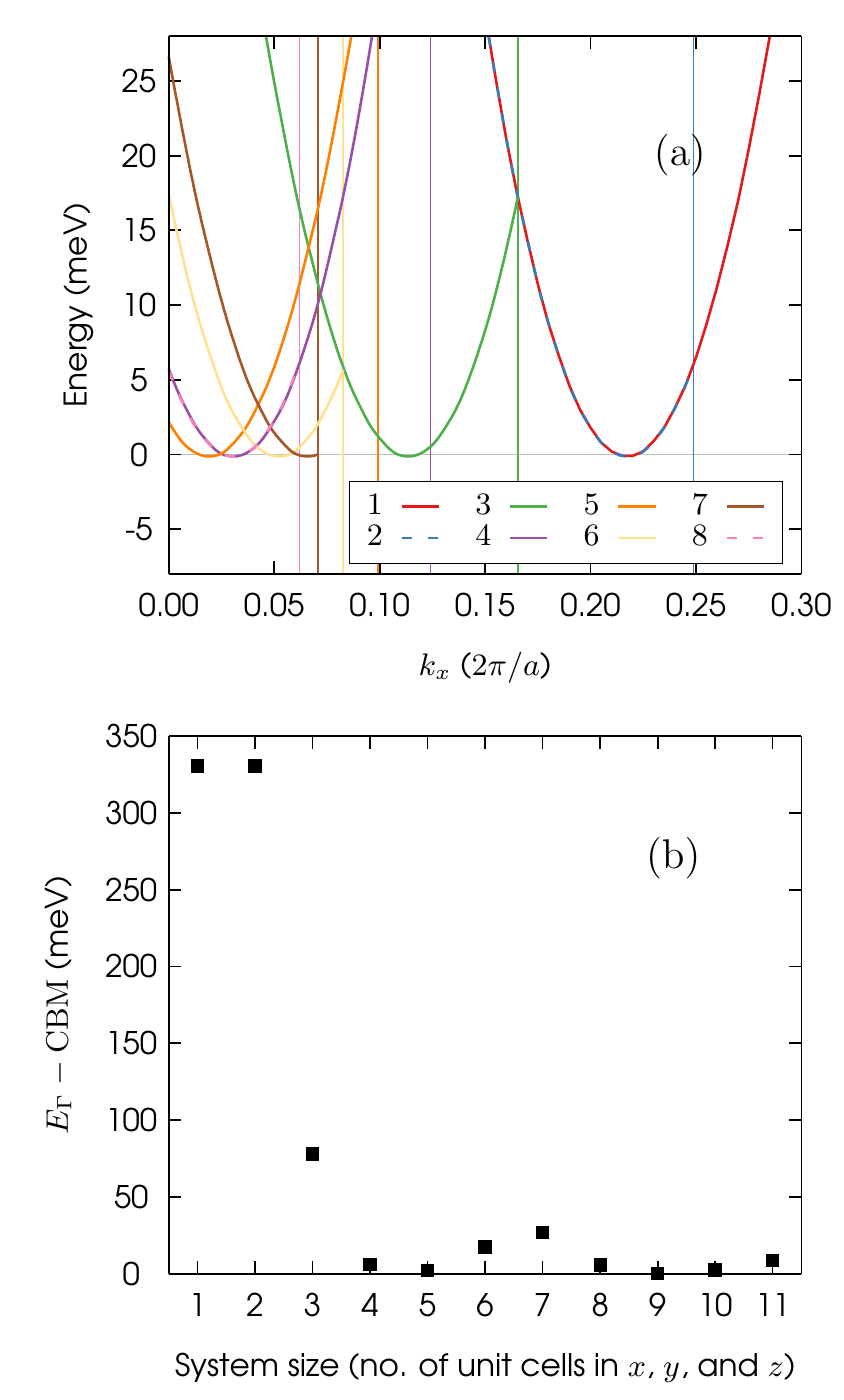}
    \caption{\label{fig:figure2}(Color online). (a) The lowest conduction valley of bulk silicon for simple cubic supercells that range in size from 8 to 4096 atoms. The key shows the dimensions of the supercells in terms of the number of simple cubic unit cells in $x$, $y$, and $z$, i.e. 8 is equivalent to $8\times8\times8$ unit cells or 4096 atoms. The boundaries of the Brillouin zones for each of the supercells are shown as vertical lines. The CBM of bulk silicon has been set to energy zero. (b) The difference between the energy of the conduction valley at the $\Gamma$ point ($E_{\Gamma}$) and the CBM for supercells that range in size from 8 to 10,648 atoms.}
\end{figure}

To reduce the computational expense of performing these electronic structure calculations, we have used a $k$~point grid that contains only a single $k$~point: the $\Gamma$ point, i.e. $k = (0,0,0)$.
For the supercell of 10,648 atoms, an increase in the size of the $k$ point grid would result in these calculations being computationally impractical.
When the number of $k$ points is increased up to $8\times8\times8$ for the supercell of 512 atoms, we find the eigenvalue of the $1s(A_{1})$ state at the $\Gamma$ point converges to a value that is approximately 5~meV greater than that of the $\Gamma$~point calculation.
The eigenvalues of the $1s(T_{2})$ and $1s(E)$ states converge to values that are approximately 1~meV greater than the result of their respective $\Gamma$~point calculations.
We expect these changes in the eigenvalues of the system to decrease as the size of the supercell is increased because the size of the corresponding Brillouin zone will decrease.
Previous calculations of an arsenic donor in silicon with DFT have also been restricted to the $\Gamma$~point~\cite{Yamamoto2009a}.

We geometrically optimised the ionic positions of supercells that ranged in size from 64 to 4096 atoms.
We found the maximum displacement of a silicon atom was largest for the supercell of 64 atoms.
When the size of the supercells is increased up to 4096 atoms the maximum displacement decreased to less than 0.02~\AA.
This displacement is equivalent to less than 0.5\% of the lattice constant of bulk silicon.
We therefore conclude it is unnecessary to relax the ionic positions of silicon atoms beyond their bulk values for supercells larger than 4096 atoms.

The conduction band minimum (CBM) of silicon is located at $|k|\approx0.85\left(2\pi / a\right)$, along each of the cardinal axes of reciprocal space, inside the face centred cubic Brillouin zone.
Because silicon is an indirect bandgap semiconductor, the energy of the lowest conduction valley at the $\Gamma$ point is not equal to the energy of the CBM.
This is a result of the dispersion of the energy bands.
We calculate the eigenvalues of the phosphorus donor electron at $|k|=0$, not $|k|\approx0.85\left(2\pi / a\right)$, and therefore it is necessary to offset the computed energies of the $1s(A_{1})$, $1s(T_{2})$, and $1s(E)$ states to find their value at $|k|\approx0.85\left(2\pi / a\right)$.

The size of the Brillouin zone is decreased when the size of the supercell is increased.
Decreasing the size of the Brillouin zone causes the bands, and therefore CBM, to be folded towards the centre of the zone, i.e. the $\Gamma$ point, in a process known as band folding~\cite{Drumm2013b}.
Consequently, the amount by which the energies of the $1s(A_{1})$, $1s(T_{2})$, and $1s(E)$ states must be offset, to account for the parabolic dispersion of the band, is different for each supercell.
The folding of the lowest conduction valley is plotted in Fig.~\ref{fig:figure2}{a} for supercells that range in size from 8 to 4096 atoms.
The value of the offset for each supercell can be computed by taking the difference between the energy of the valley at the $\Gamma$ point ($E_{\Gamma}$) and the conduction-band minimum (CBM).
These energies have been plotted for all supercells in Fig.~\ref{fig:figure2}{b}.

\begin{figure}[t!]
    \includegraphics{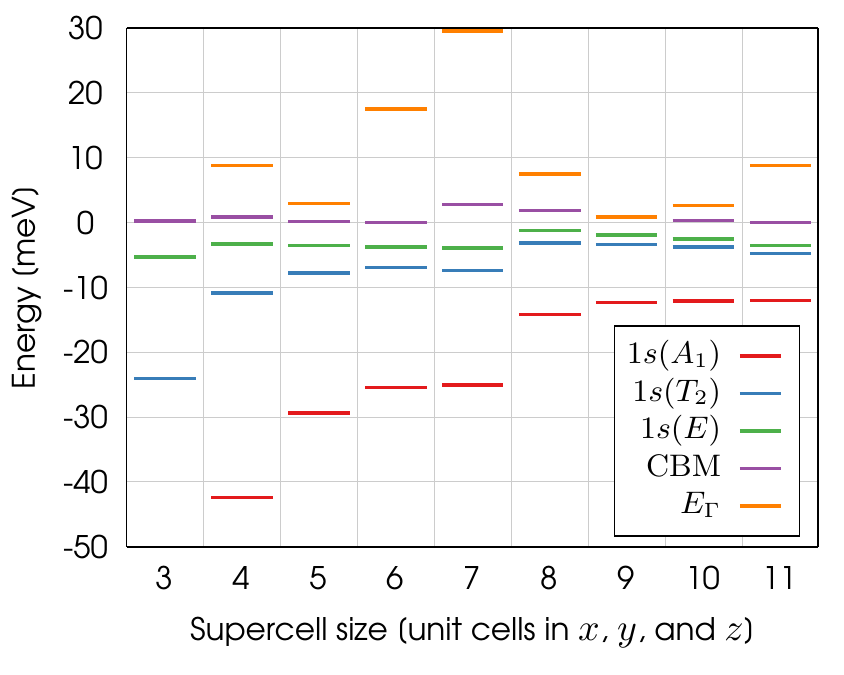}
    \caption{\label{fig:figure6}(Color online). The energy levels of the $1s(A_{1})$, $1s(T_{2})$, and $1s(E)$ states for supercells that range in size from 216 to 10,648 atoms. The energy of the CBM and the lowest conduction valley at $\Gamma$ ($E_{\Gamma}$) for bulk silicon are also shown for each supercell size. The conduction band minimum of the supercell containing 10,648 atoms has been set to energy zero.}
\end{figure}

The value of $E_{\Gamma} - \mathrm{CBM}$ decreases as the size of the supercell is increased.
As shown in Fig.~\ref{fig:figure2}{b}, this relationship is not monotonic: the CBM is not always folded closer to the $\Gamma$ point as the size of the Brillouin zone is decreased.
Figure~\ref{fig:figure2}{a} shows the lowest conduction valley for bulk silicon only.
If the dispersion of this band does not change significantly upon doping with phosphorus, then the difference $E_{\Gamma} - \mathrm{CBM}$ can be used to correct the computed energies of the $1s(A_{1})$, $1s(T_{2})$, and $1s(E)$ states.
The positions of the conduction valleys on the $k_{x}$ axis, in Fig.~\ref{fig:figure2}, have been computed by folding the band structure of bulk silicon.
The unfolded band structure was calculated using an eight atom simple cubic unit cell and a $k$ point grid of $6\times6\times6$.
For the sake of clarity, we do not show the part of the bands that are reflected back into the Brillouin zone at the zone boundary.
Neither do we show the conduction valleys of supercells with more than 4096 atoms in Fig.~\ref{fig:figure2}.
The reflection of the bands at the zone boundary is a consequence of the fact that a solution in one Brillouin zone must be a solution in all Brillouin zones~\cite{Drumm2013b}.

The energy of the lowest conduction valley of bulk silicon at the $\Gamma$ point ($E_{\Gamma}$) is shown in Fig.~\ref{fig:figure6} for every supercell studied in this work.
As expected, the value of $E_{\Gamma}$ is different for each supercell.
The conduction band minima plotted in Fig.~\ref{fig:figure6} are for bulk silicon and have been calculated by substracting $E_{\Gamma} - \mathrm{CBM}$ from $E_{\Gamma}$, i.e. $E_{\Gamma} - \left(E_{\Gamma} - \mathrm{CBM}\right) = \mathrm{CBM}$.
The CBM for bulk silicon is not expected to change as the size of the supercell is increased.
We therefore use the CBM of the supercell containing 10,648 atoms as a point of reference by setting it to energy zero in the figure.

We find the CBM for each of the supercells do not agree when the energies are corrected for band folding only.
We also need to account for the differences in the valence band maximum (VBM) of each supercell.
The conduction band minima are shifted by the difference between the VBM of each supercell and the VBM of the supercell containing 10,648 atoms.
Once this is done, the CBM in Fig.~\ref{fig:figure6} agree to within 4~meV.
The remaining discrepancies in the conduction band minima could be caused by the differing $k$ point grids that were used to calculate the quantity $E_{\Gamma} - \mathrm{CBM}$ and the conduction band minima plotted in Fig.~\ref{fig:figure6}, or similar errors in the VBM itself.
The VBM is not affected by band folding because it appears at the $\Gamma$ point in the Brillouin zone.

\subsection{\label{sec:bindingenergy}Calculation of binding energies}

In this section, we give the mathematical details of the calculation of the binding energy for the donor electron in full.
This method was first proposed in Ref.~\onlinecite{Yamamoto2009a} for an arsenic donor in silicon.

The binding energy of the donor electron is given by
\begin{equation} \label{eq:BindingEnergy}
    E = T + U
\end{equation}
where $T$ is the kinetic energy and $U$ is the potential energy of the donor electron.
In \eqref{eq:BindingEnergy} the potential energy is defined as
\begin{equation} \label{eq:PotentialEnergy}
    U = \int_{-\infty}^{\infty} \psi^{*}(\mathbf{r}) V(\mathbf{r}) \psi(\mathbf{r}) d^{3}\mathbf{r}
\end{equation}
where $\psi$ is the wavefunction of the donor electron and $V$ is the impurity potential for the phosphorus donor.
The impurity potential can be written as
\begin{equation} \label{eq:ImpurityPotential}
    V(\mathbf{r}) = V^{\mathrm{P:Si}}(\mathbf{r}) - V^{\mathrm{1e:Si}}(\mathbf{r})
\end{equation}
where $V^{\mathrm{P:Si}}$ is the electric potential for a phosphorus-doped silicon system and $V^{\mathrm{1e:Si}}$ is the electric potential for an electron-doped silicon system.
By an electron-doped silicon system, we mean a bulk silicon system with one electron added.
In contrast, for the phosphorus-doped system, one electron is added to the system by substituting a silicon atom with a phosphorus atom.
These two electric potentials can be defined as
\begin{equation} \label{eq:ElectricPotentialSiP}
    V^{\mathrm{P:Si}}(\mathbf{r}) = V_{ee}^{\mathrm{P:Si}}(\mathbf{r}) + V_{XC}^{\mathrm{P:Si}}(\mathbf{r}) + V_{eN}^{\mathrm{P:Si}}(\mathbf{r})
\end{equation}
and
\begin{equation} \label{eq:ElectricPotentialSi1e}
    V^{\mathrm{1e:Si}}(\mathbf{r}) = V_{ee}^{\mathrm{1e:Si}}(\mathbf{r}) + V_{XC}^{\mathrm{1e:Si}}(\mathbf{r}) + V_{eN}^{\mathrm{1e:Si}}(\mathbf{r})
\end{equation}
where $V_{ee}$ is the electron-electron contribution to the electric potential, $V_{XC}$ is the exchange-correlation contribution to the electric potential, and $V_{eN}$ is the electron-nuclear contribution to the electric potential.
Substituting \eqref{eq:ElectricPotentialSiP} and \eqref{eq:ElectricPotentialSi1e} into \eqref{eq:ImpurityPotential} and rearranging we have
\begin{align}
    V(\mathbf{r}) &= \left[V_{ee}^{\mathrm{P:Si}}(\mathbf{r}) + V_{XC}^{\mathrm{P:Si}}(\mathbf{r}) + V_{eN}^{\mathrm{P:Si}}(\mathbf{r})\right] - \ldots \notag \\
                      & \hspace{14mm} \ldots \left[V_{ee}^{\mathrm{1e:Si}}(\mathbf{r}) + V_{XC}^{\mathrm{1e:Si}}(\mathbf{r}) + V^{\mathrm{1e:Si}}_{eN}(\mathbf{r})\right] \notag \\
    V(\mathbf{r}) &= \left[V_{ee}^{\mathrm{P:Si}}(\mathbf{r}) - V_{ee}^{\mathrm{1e:Si}}(\mathbf{r})\right] + \left[V_{XC}^{\mathrm{P:Si}}(\mathbf{r}) - V_{XC}^{\mathrm{1e:Si}}(\mathbf{r})\right] + \ldots \notag \\
                      & \hspace{14mm} \ldots \left[V_{eN}^{\mathrm{P:Si}}(\mathbf{r}) - V^{\mathrm{1e:Si}}_{eN}(\mathbf{r})\right] \label{eq:imp_pot}
\end{align}

\begin{figure}[t!]
    \includegraphics{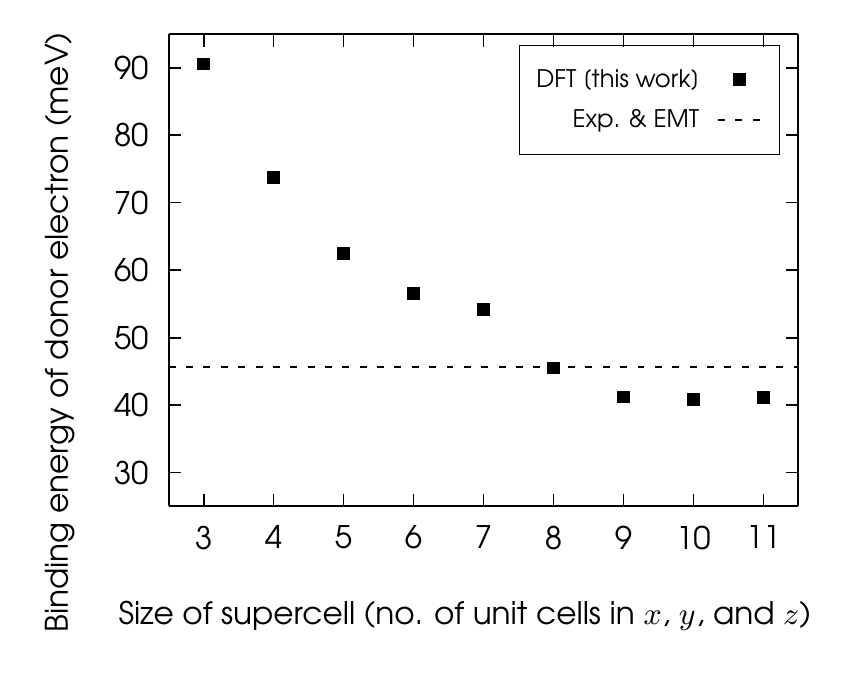}
    \caption{\label{fig:figure4}Binding energies for the $1s(A_{1})$ state of a phosphorus donor electron in silicon, calculated by the method described in this section, for supercells that range in size from 216 to 10,648 atoms. The accepted value for the binding energy, taken from Ref.~\onlinecite{Jagannath1981a}, is shown as a dashed line.}
\end{figure}

In the equations above, the impurity potential is screened by the electron-electron and exchange-correlation terms.
Next, we set $V_{ee}$ and $V_{XC}$ to zero and thereby introduce a new quantity, the unscreened impurity potential $V'$.
The unscreened impurity potential is given by the last term in \eqref{eq:imp_pot}:
\begin{equation} \label{eq:UnscreenedImpurityPotential}
    V'(\mathbf{r}) = V_{eN}^{\mathrm{P:Si}}(\mathbf{r}) - V_{eN}^{\mathrm{1e:Si}}(\mathbf{r})
\end{equation}
In our calculations, the electron-nuclear interaction is described by Troullier-Martins pseudopotentials and we can write
\begin{equation} \label{eq:PhosphorusPseudopotential}
    V_{eN}^{\mathrm{P:Si}}(\mathbf{r}) = V_{pp}^{\mathrm{P}}(\mathbf{r} - \mathbf{R}_{0}) + \sum_{i=1}^{N-1} V_{pp}^{\mathrm{Si},\mathrm{P:Si}}(\mathbf{r} - \mathbf{R}_{i})
\end{equation}
and
\begin{equation} \label{eq:ElectronPseudopotential}
    V_{eN}^{\mathrm{1e:Si}}(\mathbf{r}) = \sum_{i=0}^{N-1} V_{pp}^{\mathrm{Si},\mathrm{1e:Si}}(\mathbf{r} - \mathbf{R}_{i})
\end{equation}
where $\mathbf{R}_{0}$ is the ionic position of the phosphorus donor atom, $\mathbf{R}_{i}$ is the ionic position of silicon atom $i$, and $V_{pp}^{\mathrm{P}}$ and $V_{pp}^{\mathrm{Si}}$ are the pseudopotentials of phosphorus and silicon, respectively.
Substituting \eqref{eq:PhosphorusPseudopotential} and \eqref{eq:ElectronPseudopotential} into \eqref{eq:UnscreenedImpurityPotential}, we obtain
\begin{equation*}
    V'(\mathbf{r}) = V_{pp}^{\mathrm{P}}(\mathbf{r} - \mathbf{R}_{0}) + \sum_{i=1}^{N-1} V_{pp}^{\mathrm{Si},\mathrm{P:Si}}(\mathbf{r} - \mathbf{R}_{i}) - \sum_{i=0}^{N-1} V_{pp}^{\mathrm{Si},\mathrm{1e:Si}}(\mathbf{r} - \mathbf{R}_{i})
\end{equation*}
which, because we have not relaxed the ionic positions of the silicon atoms after phosphorus substitution, simplifies to
\begin{equation*}
    V'(\mathbf{r}) \approx V_{pp}^{\mathrm{P}}(\mathbf{r} - \mathbf{R}_{0}) - V_{pp}^{\mathrm{Si}}(\mathbf{r} - \mathbf{R}_{0})
\end{equation*}
where $V_{pp}^{\mathrm{Si}} \equiv V_{pp}^{\mathrm{Si},\mathrm{P:Si}} \approx V_{pp}^{\mathrm{Si},\mathrm{1e:Si}}$ is approximate because the norm-consering Troullier-Martins pseudopotentials are nonlocal.
Let $\mathbf{R}_{0} = (0, 0, 0)$, then
\begin{equation} \label{eq:unscreenedImpurityPotential}
    V'(\mathbf{r}) \approx V_{pp}^{\mathrm{P}}(\mathbf{r}) - V_{pp}^{\mathrm{Si}}(\mathbf{r})
\end{equation}
That is, the unscreened impurity potential is given by the difference in the pseudopotentials for phosphorus and silicon.
We have used only the $l=0$ component of the norm-conserving pseudopotentials for phosphorus and silicon when evaluating Eq.~\ref{eq:unscreenedImpurityPotential}.
This approximation is justified given the structure of the eigenfunction for the $1s(A_{1})$ state (cf. Fig.~1 in the main text).
Electron screening can now be reintroduced using the following description.
We rewrite the screened impurity potential as~\cite{Yamamoto2009a}
\begin{equation} \label{eq:ScreenedImpurityPotential}
    V(\mathbf{r}) = \int_{-\infty}^{\infty} \epsilon^{-1}(q) V'(q) \exp{(-i \mathbf{q} \cdot \mathbf{r})} \frac{d^{3}q}{\left(2\pi\right)^{3}}
\end{equation}
where $V'(q)$ is the Fourier transform of the unscreened impurity potential. The dielectric screening is described by a nonlinear function~\cite{Pantelides1974a,Nara1966a}
\begin{equation} \label{eq:dielectricfunction}
    \epsilon^{-1}(q) = \frac{A q^{2}}{q^{2} + \alpha^{2}} + \frac{\left(1 - A\right) q^{2}}{q^{2} + \beta^{2}} + \frac{\gamma^{2}}{\epsilon(0)\left(q^{2} + \gamma^{2}\right)}
\end{equation}
with $A = 1.175$, $\alpha = 0.7572$, $\beta = 0.3123$, $\gamma = 2.044$, and $\epsilon(0) = 11.4$.
The constants $A$, $\alpha$, $\beta$, and $\gamma$ were found by fitting the above function to the $q$ dependent dielectric screening in silicon, which was calculated from the random phase approximation~\cite{Pantelides1974a}.
We can then use \eqref{eq:ScreenedImpurityPotential} to calculate the potential energy of the donor electron using \eqref{eq:PotentialEnergy}.
Finally, to calculate the kinetic energy of the donor electron, we use the virial theorem:
\begin{equation} \label{eq:KineticEnergy}
    T = \frac{1}{2} \int_{-\infty}^{\infty} \psi^{*}(\mathbf{r}) \left(\frac{dV(\mathbf{r})}{d\mathbf{r}} \cdot \mathbf{r}\right) \psi(\mathbf{r}) d^{3}\mathbf{r}
\end{equation}
The binding energy of the donor electron can then be calculated from the kinetic and potential energies using~\eqref{eq:BindingEnergy}.

The binding energies of the donor electron's ground state, calculated using supercells of 216 to 10,648 atoms, are shown in Fig.~\ref{fig:figure4}.
For the supercell of 10,648 atoms, the value of the binding energy is converged to within 1~meV.
The accepted value for the binding energy of the donor electron's ground state, which is equal to 45.59~meV, is shown as a dashed line in Fig.~\ref{fig:figure4}.
The supercell of 216 atoms overestimates the binding energy of the $1s(A_{1})$ state in the figure but the binding energy decreases as the size of the supercell is increased.
This energy is within 5~meV of the accepted value for a supercell of 10,648 atoms.
Unfortunately there is no systematic way of calculating the uncertainty in this energy, but it is unlikely that the uncertainty is less than 5~meV.

\bibliography{articles-SDp1.bib}

\end{document}